\documentclass[english,aps,prb,twocolumn,amsmath,amssymb,showpacs,superscriptaddress,notitlepage,longbibliography]{revtex4-2}
\usepackage{bm}
\usepackage[pdftex]{graphicx,hyperref}
\hypersetup{colorlinks = true, urlcolor = blue, linkcolor = blue, citecolor = blue}
\usepackage{mathtools}
\usepackage{ulem}

\begin{document}

\title{Van Hove singularity-induced multiple magnetic transitions in multi-orbital systems}

\author{Chen Lu}
\email{luchen@hznu.edu.cn}
\affiliation{School of Physics and Hangzhou Key Laboratory of Quantum Matter, Hangzhou Normal University, Hangzhou 311121, China}
\author{Lun-Hui Hu}
\email{lunhui@zju.edu.cn}
\affiliation{Center for Correlated Matter and School of Physics, Zhejiang University, Hangzhou 310058, China}

\begin{abstract}
Van Hove singularities (VHSs) amplify electronic correlations, providing a crucial platform for discovering novel quantum phase transitions. Here, we show that VHSs in multi‑orbital systems can stabilize a variety of competing $\bm{Q}=0$ magnetic orders, including intrinsic altermagnetism emerging from spontaneous orbital antiferromagnetism. This intrinsic phase, in which antiparallel spins reside on distinct orbitals, is realized across all four 2D Bravais lattices. It is driven by orbital‑resolved spin fluctuations enhanced by inter‑orbital hopping and favors suppressed Hund's coupling $J_H$, strong inter‑orbital hybridization, and filling near a VHS from quadratic band touching. Through Hubbard‑$U$–$J_H$ phase diagrams we map several magnetic phase transitions: (i) ferrimagnet to $d$‑wave extrinsic altermagnet, (ii) $d$‑wave intrinsic altermagnet to ferromagnet, and (iii) $g$‑wave extrinsic altermagnet to either $d$‑wave extrinsic altermagnet or ferromagnet. Our work identifies VHSs as a generic route to altermagnetism in correlated materials.
\end{abstract}

\maketitle

\noindent{\bf Introduction}\\
Van Hove singularities (VHSs) originate from saddle points in the electronic band structure and typically produce a logarithmic divergence in the density of states~\cite{van1953occurrence}. When the Fermi level lies near such a singularity, electronic correlations are strongly enhanced, making the system a fertile platform for exploring symmetry-breaking orders, including unconventional superconductivity~\cite{kohn1965new,honerkamp2001magnetic,piriou2011first,hu2022rich,park2021tunable,dzyaloshinskii1987superconducting,markiewicz1991van,radtke1994relation,newns1991role,mcchesney2010extended,yao2015topological}, charge density waves~\cite{rice1975new,lin2021complex,wu2023pair,gonzalez2001charge,cho2021emergence}, and magnetism~\cite{gonzalez2000kinematics,fleck1997magnetic,liu2019magnetism,yin2019negative,vozmediano2002properties,kampf2003competing,igoshev2007magnetic,gonzalez2013magnetic,wang2013competing,ziletti2015van,goh2016mechanism,wang2018electronic,sherkunov2018electronic,liu2019magnetism,wu2024discovery,park2025spin}. In multi-orbital systems, the influence of VHSs is further enriched by orbital-selective saddle points, which can generate complex, competing magnetic orders. Thus, tuning microscopic parameters (e.g., doping, strain) can induce a redistribution of spin polarization across orbitals or sublattices, often favoring exotic compensated magnetic states over conventional ferromagnetic or N\'eel antiferromagnetic orders~\cite{yu2025altermagnetism,giuli2025altermagnetism,bose2024altermagnetism,li2012spontaneous}.

One prominent example of unconventional magnetism is altermagnetism, a recently identified collinear antiferromagnetic phase characterized by vanishing net magnetization and momentum-dependent spin-split electronic bands~\cite{noda2016momentum,naka2019spin,ahn2019antiferromagnetism,hayami2019momentum,vsmejkal2020crystal,yuan2020giant,shao2021spin,mazin2021prediction,ma2021multifunctional,yuan2021prm,vsmejkal2022beyond}. The altermagnetic phase has been experimentally confirmed in diverse materials~\cite{feng2022anomalous,fedchenko2024ruo2,lin2024ruo2,gonzalez2023MnTe,krempasky2024MnTe,lee2024MnTe,osumi2024MnTe,liu2024chiral,reimers2024CrSb,ding2024CrSb,yang2025three,zhang2025crystal,jiang2024discovery,jiang2025metallic,xu2025electronic,yang2025interface,zhang2025x}, and provides promising platforms for novel spintronic functionalities~\cite{vsmejkal2022emerging,gonzalez2021efficient,vsmejkal2022giant,guo2024emerging,shao2024antiferromagnetic,song2025altermagnets}. Alternative theoretical mechanisms for non-relativistic spin-splitting, such as spin-channel Pomeranchuk instabilities~\cite{Hirsch1990prb,wu2004dynamic,wu2007fermi} and $d$-wave spin-density waves~\cite{Ikeda1998prl}, have also been proposed. In altermagnets, the compensated spin sublattices feature antiparallel spins that can be only interchanged via crystalline symmetries, such as rotation and reflection~\cite{vsmejkal2022emerging}, resulting in the breaking of both composite ${\cal P}\cdot{\cal T}$ (parity-time reversal) and $t\cdot{\cal T}$ (translation-time reversal) symmetries~\cite{bai2024altermagnetism,fender2025altermagnetism}. These characteristics give rise to various momentum-space spin textures, including $d$-, $g$-, and $i$-wave textures~\cite{Fernandes2024prb}.

From a symmetry perspective, altermagnetism represents a distinct class of correlated phases~\cite{liu2025different}. Given its considerable potential for antiferromagnetic spintronics, numerous theoretical mechanisms for realizing altermagnetism have been proposed, including orbital ordering~\cite{leeb2024spontaneous}, stacking twisted van der Waals ferromagnets~\cite{He2023prl,liu2024twisted}, coupling to ferroelectrics~\cite{vsmejkal2024altermagnetic,duan2025antiferroelectric,gu2025ferroelectric}, and engineering lattice vacancies~\cite{chakraborty2024strain,zhu2025design,li2025pressure}. While correlation effects are recognized as central to magnetism~\cite{maier2023weak,Roig2024prb,das2024realizing,Sato2024prl,Zhao2025prb}, the role of VHSs in driving transitions between altermagnetic and other collinear orders remains largely unexplored.

In this work, we investigate the conditions under which VHSs stabilize altermagnetic phases. Using analytical arguments and random-phase-approximation calculations, we show that inter-orbital hopping strongly enhances orbital-resolved spin fluctuations, which in turn stabilizes an altermagnetic phase emerging from orbital antiferromagnetism. This phase is favored under conditions of large Hubbard repulsion $U$ but small Hund's coupling $J_H$. By computing the staggered susceptibilities for various collinear magnetic phases, we construct comprehensive $U$-$J_H$ phase diagrams. These diagrams reveal rich phase competition, including transitions between extrinsic/intrinsic altermagnetism, ferromagnetism, ferrimagnetism, and N\'eel antiferromagnetism. Moreover, our work establishes that VHSs arising from quadratic Dirac band touching can play a central mechanism for stabilizing altermagnetism.

\vspace{0.5\baselineskip}
\noindent{\bf Symmetry classification and prototype models} \\ 
We begin with a general symmetry analysis to establish orbital antiferromagnetism as a mechanism for altermagnetism in two-dimensional (2D) systems. To illustrate this, we consider a 2D Bravais lattice with a single atom per site hosting two locally degenerate (or nearly degenerate) orbitals, such as $\{ p_x, p_y\}$ or $\{ d_{xz}, d_{yz}\}$. The relevant point-group symmetries are $C_2$ for oblique lattice, $C_{2v}$ for rectangular and centered rectangular lattice, $C_{4v}$ for square lattice, and $C_{6v}$ for hexagonal lattice. Inversion ${\cal P}$ in this case is equivalent to $C_{2z}$. The combination of these point groups with time-reversal symmetry ${\cal T}$ governs the symmetry-breaking patterns of orbital antiferromagnetic order, and we define the orders as 
\begin{align}
\label{eq-order1}
\hat{\cal O}_1(\bm{r}) &= c_{p_x,\uparrow}^\dagger(\bm{r}) c_{p_x,\uparrow}(\bm{r}) - c_{p_x,\downarrow}^\dagger(\bm{r}) c_{p_x,\downarrow}(\bm{r}) \nonumber \\
&- c_{p_y,\uparrow}^\dagger(\bm{r}) c_{p_y,\uparrow}(\bm{r}) + c_{p_y,\downarrow}^\dagger(\bm{r}) c_{p_y,\downarrow}(\bm{r}), \\
\label{eq-order2}
\hat{\cal O}_2(\bm{r}) &= c_{p_+,\uparrow}^\dagger(\bm{r}) c_{p_+,\uparrow}(\bm{r}) - c_{p_+,\downarrow}^\dagger(\bm{r}) c_{p_+,\downarrow}(\bm{r}) \nonumber \\
&- c_{p_-,\uparrow}^\dagger(\bm{r}) c_{p_-,\uparrow}(\bm{r}) + c_{p_-,\downarrow}^\dagger(\bm{r}) c_{p_-,\downarrow}(\bm{r}),
\end{align}
where $\bm{r}$ denotes the lattice site and $c_{p_{\pm}}=(c_{p_x} \pm c_{p_y})/\sqrt{2}$ are the electron creation operators. Both $\hat{\cal O}_1(\bm{r})$ and $\hat{\cal O}_2(\bm{r})$ encode spin-orbital correlations with opposite spin polarization on distinct orbitals. We assume a long-range order such that $\langle \hat{\cal O}(\bm{r}) \hat{\cal O}(0) \rangle$ remains finite as $r\to\infty$, and further that the order parameter $\langle \hat{\cal O}(\bm{r}) \rangle$ preserves the translational symmetry of the unit cell (i.e.,~$t\cdot{\cal T}$ is broken). Under these conditions, the orders $\langle \hat{\cal O}(\bm{r}) \rangle$ can be classified according to the groups $C_2\otimes {\cal T}$, $C_{2v}\otimes {\cal T}$, $C_{4v}\otimes {\cal T}$, and $C_{6v}\otimes {\cal T}$ groups. In the absence of spin-orbit coupling, the symmetry of the system is described by a spin-space group. In this description, a symmetry operation $\{g_s||g_r\}$ acts independently on real space ($g_r$) and spin space ($g_s$). For example: the group $C_2\otimes {\cal T}$ is generated by $\{E||C_{2z}\}$ and $\{{\cal T}||C_{2z}\}$, where ${\cal T}$ is the spinful time-reversal operator satisfying ${\cal T}^2=-1$.

Our symmetry classification, summarized in Table~\ref{table1}, reveals that the $\{{\cal T}||C_{2z}\}$ symmetry (or equivalently ${\cal P}\cdot{\cal T}$) is broken in all cases, giving rise to the characteristic spin-splitting of altermagnetism. However, altermagnetism further requires a vanishing net magnetization, a condition that leads to key differences among lattice types. For any symmetry operator $g$, its action on the order parameter in Eqs.~\eqref{eq-order1} and \eqref{eq-order2} follows $g \hat{\cal O}_{1,2} g^{-1}$. When the order parameter respects the symmetry, $g \hat{\cal O}_{1,2} g^{-1} = \hat{\cal O}_{1,2}$; conversely, if the operation changes the order parameter, the symmetry is broken, as indicated by the $-$ or $\mathcal{\times}$ entries in Table~\ref{table1}. In practice, the oblique lattice cannot host altermagnetism because its antiferromagnetic order lacks the symmetry constraints needed to enforce zero net magnetization. By contrast, the ${\cal O}_2$ phase preserves $\{{\cal T}||C_{2x}\}$ symmetry in rectangular, centered rectangular, and square lattices, ensuring full spin compensation and realizing $d_{x^2-y^2}$-wave altermagnetism. Meanwhile, the ${\cal O}_1$ order produces $d_{x^2-y^2}$-wave altermagnetism only in square lattices, which will be addressed later with explicit model calculations. Because both $\mathcal{O}_1$ and $\mathcal{O}_2$ spontaneously break both ${\cal P}\cdot{\cal T}$ and $t\cdot{\cal T}$ symmetries, we refer to them as \textit{intrinsic altermagnetic phase}. Moreover, in hexagonal lattices, ${\cal O}_2$ explicitly breaks $C_{6z}$ symmetry, giving rise to a nematic altermagnetic state~\cite{camerano2025multiferroic,li2025submit}.

\begin{table}[t]
\centering
\vspace{-0.6em}
\caption{Symmetry classification of the ${\cal O}_{1/2}$ orbital-AFM orders (defined in Eqs.~\eqref{eq-order1} and \eqref{eq-order2}) across five Bravais lattices.
(AM = altermagnet; AFM = antiferromagnet)
}
\includegraphics[width=\linewidth]{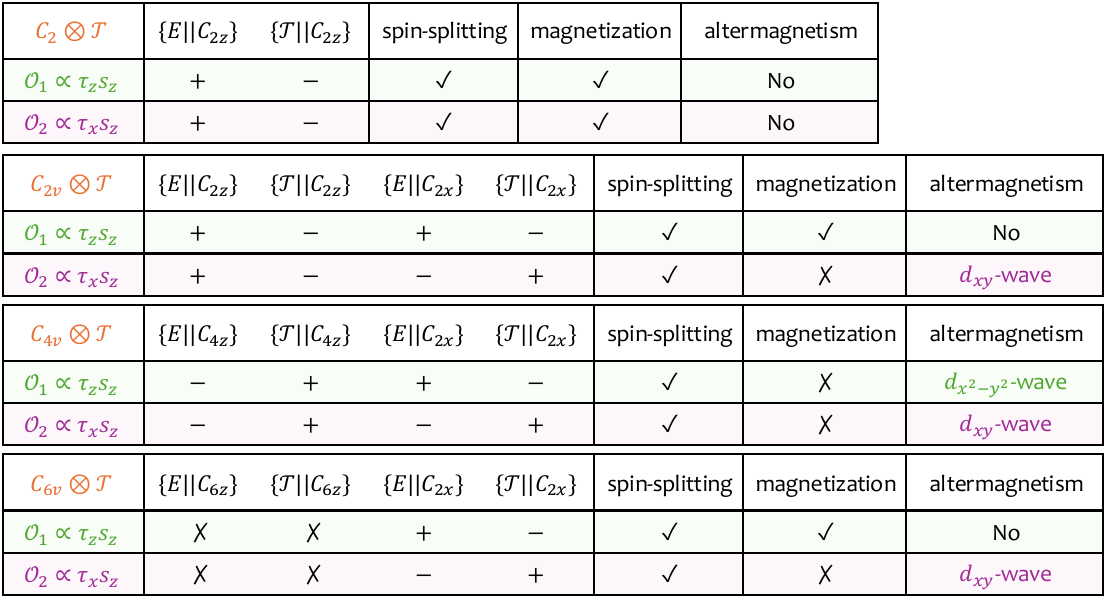}
\label{table1}
\end{table}

\begin{figure}[t]
\centering
\includegraphics[width=0.48\textwidth]{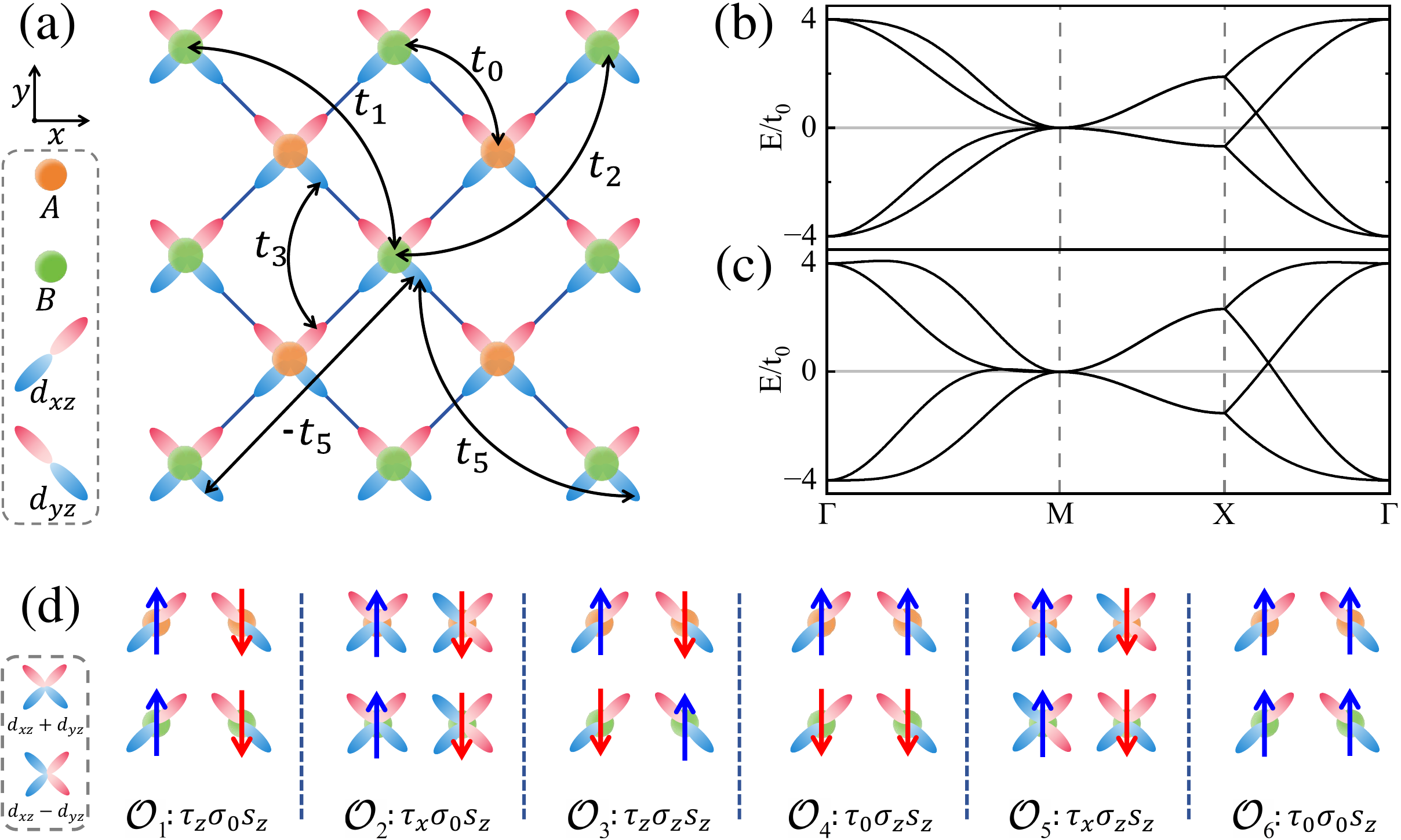}
\caption{Competing magnetic orders in a two-orbital square lattice model.
(a) Schematic of the square lattice with two orbitals ({$d_{xz}$, $d_{yz}$}) and $A$ (orange) and $B$ (green) sublattices. Hopping parameters include intra-orbital terms ($t_0, t_1, t_2$) and and inter-orbital terms ($t_3, t_5$).
(b) Tight-binding band structure with parameters $\{t_0,t_1,t_2,t_3,t_5\}=\{ 1,0.05,0.05,0.48,0.29\}$. Gray line is $\mu=-0.19$.
(c) Tight-binding band structure with parameters $\{t_0,t_1,t_2,t_3,t_5\}=\{ 1,0.1,0.05,0.32,0.13\}$. Gray line is $\mu=-0.3$.
(d) Six possible $\bm{Q}=\bm{0}$ magnetic order configurations within a unit cell, where blue ($\color{blue}\uparrow$) and red ($\color{red}\downarrow$) arrows represent the spin polarization of magnetic moments.
}
\label{Lattice}
\end{figure}

While the symmetry-adapted analysis has focused primarily on orbital antiferromagnetism, a complete treatment must also incorporate other competing magnetic orders. These include the $\mathbf{Q}=(\pi,\pi)$ N\'eel order~\footnote{This corresponds to the $\mathbf{Q}=0$ magnetic order in use of unit cell gauge formed by combining the two sublattices.} (relevant for both N\'eel antiferromagnetism and extrinsic altermagnetism), ferromagnetic order, and ferrimagnetic order. These distinct phases typically emerge in different parameter regimes of correlated systems. To systematically identify altermagnetism in competition with these phases, we study the minimal two-orbital Hubbard model on a square lattice. As illustrated in Fig.~\ref{Lattice}(a), the system contains two sublattices (labeled with $A$ and $B$) per unit cell, each hosting two degenerate orbitals (e.g.,~$d_{xz}$ and $d_{yz}$). Our Hamiltonian treats both the limiting case of equivalent sublattices ($A = B$) and the general case with broken sublattice-inversion symmetry. The tight-binding model is given by
\begin{align}\label{Hamiltonian_tb}
\begin{split}
H_{tb}(\bm{k}) &= \tau_0\left [ \epsilon_0(\bm{k}) \sigma_0 + \text{Re}\left [\epsilon_1(\bm{k}) \right ]\sigma_x     -\text{Im}\left [\epsilon_1(\bm{k}) \right ]\sigma_y \right.  \\
&\left. + \epsilon_3(\bm{k}) \sigma_z \right ] + \left [ f_1(\bm{k})\tau_x +f_3(\bm{k})\tau_z \right ]\sigma_0,
\end{split}
\end{align}
where $\epsilon_0(\bm{k})=-(t_1+t_2)[\cos(k_x+k_y)+\cos(k_x-k_y)]-\mu$ with $\mu$ the chemical potential, $\epsilon_1(\bm{k})=-t_0[1+ e^{ik_x} + e^{ik_y} + e^{i(k_x+k_y)}]$, $\epsilon_3(\bm{k})=(t_2-t_1)[\cos(k_x+k_y)-\cos(k_x-k_y)]$, $f_1(\bm{k})=-2t_3(\text{cos}\, k_x-\text{cos}\, k_y)$, $f_3(\bm{k})=2t_5[\cos( k_x+k_y)-\cos( k_x-k_y)]$, and $\tau_{x,y,z}$ and $\sigma_{x,y,z}$ are Pauli matrices acting on the orbital and sublattice degrees of freedom, respectively. These hopping parameters are illustrated in Fig.~\ref{Lattice}(a). Among them, $\{t_0,t_1,t_2\}$ are orbital-independent, while $\{t_3,t_5\}$ are orbital-dependent. We set $t_0=1$ as the energy unit of the system. In this work, we consider both the case where $t_1 = t_2$ (corresponding to identical $A$ and $B$ sublattices, i.e., the $1\times1$ square lattice) and another case with $t_1\neq t_2$ (i.e., the $\sqrt{2}\times\sqrt{2}$ square lattice). Both cases are crucial for the emergence of altermagnetic phases. For instance, the $\mathbf{Q}=(\pi,\pi)$ N\'eel antiferromagnetic order yields altermagnetic spin-splitting bands only in the $t_1\neq t_2$ case~\cite{maier2023weak}. We refer to this as \textit{extrinsic altermagnetic phase}. This mechanism, in fact, describes the majority of recently discovered altermagnetic materials~\cite{jungwirth2026symmetry}.

In the non-interacting case, Fig.~\ref{Lattice}(b) and (c) show the band structures for $t_1 = t_2$ and $t_1 \ne t_2$, respectively. In both scenarios, the chemical potential is aligned near VHS at the $\textbf{M}$ point. This choice enables a direct examination of correlation effects. The repulsive Hubbard-Hund interaction is,
\begin{align}\label{Hamiltonian}
\begin{split}
H_{int}&=U\sum _{\bm{i},\tau} n_{\bm{i}\tau \uparrow }n_{\bm{i}\tau \downarrow }+V\sum_{\bm{i}s,s^{\prime}} n_{\bm{i},x,s}n_{\bm{i},y,s^{\prime}} \\
&+J_H \sum _{\bm{i}}  \sum _{s,s^{\prime}} c^{\dagger}_{\bm{i},x,s}c^{\dagger}_{\bm{i},y,s^{\prime}}c_{\bm{i},x,s^{\prime}}c_{\bm{i},y,s}  \\
&+J_H \sum _{\bm{i}} c^{\dagger}_{\bm{i},x,\uparrow} c^{\dagger}_{\bm{i},x,\downarrow} c_{\bm{i},y,\downarrow} c_{\bm{i},y,\uparrow} + h.c.,
\end{split}
\end{align}
where $c_{\bm{i},\tau,s}$ is the electron annihilation operator at site $\bm{i}$ with orbital $\tau$ and spin $s$, $n_{i\tau s}=c_{\bm{i},\tau,s}^\dagger c_{\bm{i},\tau,s}$ is the corresponding density operator, $\tau=\{x,y\}$ labels the $\{d_{xz}, d_{yz}\}$ orbitals, and $s=\{\uparrow,\downarrow\}$ denotes the spin. Here, $U$ is the intra-orbital Hubbard interaction, $V$ is the inter-orbital Hubbard term, and $J_H$ is the Hund's coupling. The spin rotation symmetry imposes the constraint $U=V+2J_H$ \cite{castellani1978magnetic}. Then, we treat $U$ and $J_H/U$ as tuning parameters to explore both \textit{intrinsic} and \textit{extrinsic} altermagnetic phases, as well as their transitions to other magnetic orders.

Based on Eq.~\eqref{Hamiltonian}, one can find that $\mathcal{O}_1$ originates from the mean-field decomposition of both $U$ and $V$ terms, whereas $\mathcal{O}_2$ emerges from the $J_H$ term. Their matrix representations are given by 
\begin{align} \label{eq-op-o1o2}
\mathcal{O}_1 \triangleq \tau_z\sigma_0 s_z 
\text{ and }     
\mathcal{O}_2 \triangleq \tau_x\sigma_0 s_z,
\end{align}
where $s_z$ denotes the spin Pauli matrix. Furthermore, the $U$-$V$-$J_H$ interactions permit three collinear-type staggered orders ($\mathcal{O}_{3,4,5}$) and ferromagnetic order ($\mathcal{O}_6$),
\begin{align} \label{eq-op-o3o6}
\begin{split}
&\mathcal{O}_3 \triangleq \tau_z\sigma_z s_z
\text{ and }
\mathcal{O}_4 \triangleq \tau_0\sigma_z s_z, \\
&\mathcal{O}_5 \triangleq \tau_x\sigma_z s_z
\text{ and }
\mathcal{O}_6 \triangleq \tau_0\sigma_0 s_z.
\end{split}    
\end{align}
In this work, we focus on both \textit{intrinsic} and \textit{extrinsic} altermagnetic phases, where $\bm{Q}=\bm{0}$ serves as a necessary (though not sufficient) condition. This constraint limits the competing orders to the ${\cal O}_{1\to6}$ phases. The corresponding real-space spin configuration, displayed in Fig.~\ref{Lattice}(d), features antiparallel magnetic moments represented by blue ($\color{blue}\uparrow$) and red ($\color{red}\downarrow$) arrows.

\begin{table}[t]
\centering
\vspace{-1em}
\caption{Summary of the six $\bm{Q}=0$ magnetic phases ${\cal O}_1$–${\cal O}_6$ defined in the unit-cell basis corresponding to the $\sqrt{2}\times\sqrt{2}$ lattice. Abbreviations: iAM = intrinsic altermagnetism; AFM = antiferromagnetism; FerriM = ferrimagnetism; eAM = extrinsic altermagnetism; FM = ferromagnetism.}
\begin{tabular}{c|c|c|c|c|c|c}
\hline\hline
orders & ${\cal O}_1$ & ${\cal O}_2$ & ${\cal O}_3$ & ${\cal O}_4$ & ${\cal O}_5$ & ${\cal O}_6$ \\ \hline
$t_1=t_2$       & iAM  & iAM & AFM    & AFM          & AFM & FM  \\
$t_1\neq t_2$   & iAM  & iAM & FerriM & $d$-wave eAM & $g$-wave eAM & FM  \\ \hline\hline
\end{tabular}
\label{table2}
\end{table}

Before proceeding with calculations, we present a symmetry analysis of the staggered orders (${\cal O}_{1\to 5}$). When $t_1= t_2$, the $1 \times 1$ square lattice preserves the $C_{4v}$ point group symmetry. Based on Table~\ref{table1}, both ${\cal O}_1$ and ${\cal O}_2$ correspond to intrinsic $d$-wave altermagnetism. In contrast, the remaining three orders (${\cal O}_{3/4/5}$) involve the $\sigma_z$ matrix, which spontaneously enlarges the magnetic unit cell from $1\times 1$ to $\sqrt{2}\times\sqrt{2}$, analogous to the conventional N\'eel order. Owing to the combined parity-time symmetry ${\cal P}\cdot{\cal T}$, where ${\cal P}$ swaps sublattices A and B, these three orders \textit{cannot} produce spin-split bands. Consequently, ${\cal O}_{1/2}$ and ${\cal O}_{3/4/5}$ belong to distinct symmetry representations. However, ${\cal P}$ can be broken in the non-interacting Hamiltonian, for instance, by setting $t_1 \ne t_2$, as used in this work. Such a term can originate from non-magnetic atoms~\cite{maier2023weak}, lattice strain~\cite{chakraborty2024strain} or Jahn-Teller distortion~\cite{wang2024submit}, thereby enlarging the crystalline unit cell to $\sqrt{2}\times\sqrt{2}$. This breaking of ${\cal P}$ enables extrinsic spin-splitting bands in ${\cal O}_{3/4/5}$ phases. In particular, both ${\cal O}_{4}$ and ${\cal O}_{5}$ exhibit zero net magnetization and realize extrinsic altermagnetism, whereas ${\cal O}_{3}$ breaks all the symmetries required for antiferromagnetic classification and instead shows characteristics of a ferrimagnetic state. We summarize these analysis in Table~\ref{table2}.

\vspace{0.5\baselineskip}
\noindent{\bf Symmetry-breaking magnetic instabilities} \\
To analyze magnetic instabilities, we employ the standard multi-orbital random-phase approximation (RPA) approach~\cite{scalapino1986d,moriya2012spin,hamann1969properties,zhang2025spin}. RPA has been extensively applied to conventional N\'eel and ferromagnetic orders. The bare susceptibility tensor in momentum space is defined as,
\begin{align}\label{chi0}
[\chi^{(0)}(\bm{k},\tau)]_{l_3l_4}^{l_1l_2} &\equiv \frac{1}{N}\sum_{\bm{k}_1\bm{k}_2}  \langle T_\tau c^{\dagger}_{l_1}(\bm{k}_1,\tau)c_{l_2}(\bm{k}_1+\bm{k},\tau)  \notag\\
&\times  c^{\dagger}_{l_3}(\bm{k}_2+\bm{k},0)c_{l_4}(\bm{k}_2,0)  \rangle _0,
\end{align}
where $\langle \dots \rangle_0$ denotes the thermal average for the non-interacting system, $N$ is the lattice size, $\tau$ is imaginary time, $T_\tau$ is the imaginary time-ordering operator, and ${l_1}, {l_2}, {l_3}, {l_4} \in\left \{ 1, 2, 3, 4 \right \} $ are sublattice-orbital indices corresponding to the basis states $\{ \vert A,d_{xz} \rangle, \vert A,d_{yz} \rangle, \vert B,d_{xz} \rangle, \vert B,d_{yz} \rangle \}$. The complete computational formulation for Eq.~\eqref{chi0} is provided in Methods, where we perform Fourier transformation on $\tau$ and work on the Matsubara frequency ($i\omega_n$) domain. The spin-orbit coupling is excluded in this work, and consequently, the electron operator in Eq.~\eqref{chi0} is spin-independent (i.e., carries no explicit spin index). We emphasize that the Hamiltonian in Eq.~\eqref{Hamiltonian_tb} is expressed in the unit-cell gauge. Therefore, $[\chi^{(0)}(\bm{k},i\omega_n)]_{l_1l_2}^{l_3l_4}$ having its maximum at $\bm{k}\neq0$ indicates breaking of the unit-cell translational symmetry. However, the magnetic orders $\mathcal{O}_{1\to6}$ defined above preserve this translational symmetry and consequently must appear at $\bm{k}=0$. All relevant information can be extracted from the bare susceptibility tensor $[\chi^{(0)}(\bm{k},i\omega_n)]_{l_3l_4}^{l_1l_2}$. In particular, the static bare susceptibilities for $\mathcal{O}_{1\to6}$ are given by
\begin{align}\label{chiO}
\chi^{(0)}_{\alpha}(\bm{k}) = \frac{1}{2} \sum_{l_1l_2l_3l_4}  [\bar{\cal O}_\alpha]_{l_1l_2} [\bar{\cal O}_\alpha]_{l_3l_4} [\chi^{(0)} ]^{l_1l_2}_{l_3l_4},
\end{align}
where $\chi^{(0)} \equiv \chi^{(0)}(\bm{k},i\omega_n=0)$ and $\bar{\cal O}_{\alpha}$ encode the orbital-sublattice components of the corresponding order parameters ${\cal O}_{\alpha}$ [Eqs.~\eqref{eq-op-o1o2} and \eqref{eq-op-o3o6}], given by: $\bar{\cal O}_{1\to6} = \{ \tau_z \sigma_0, \tau_x\sigma_0, \tau_z\sigma_z, \tau_0\sigma_z, \tau_x\sigma_z, \tau_0\sigma_0 \}$. To realize the intrinsic altermagnetism (${\cal O}_{1/2}$), we scan the band parameter space within physically plausible limits. In Sec.~A of Supplementary Material (SM), we present $\chi^{(0)}_{\alpha}(\bm{k})$ along high-symmetry lines and identify the leading magnetic instability. For the band parameters adopted in Fig.~\ref{Lattice}(b), the ${\cal O}_1$ (intrinsic altermagnetism) order dominates, whereas for those in Fig.~\ref{Lattice}(c), the ${\cal O}_3$ (ferrimagnetism) order becomes the most relevant instability. In addition, intrinsic altermagnetic order is obtained over a wide range of parameters [see Sec.~B of SM].

\begin{figure}[t]
\centering
\includegraphics[width=0.48\textwidth]{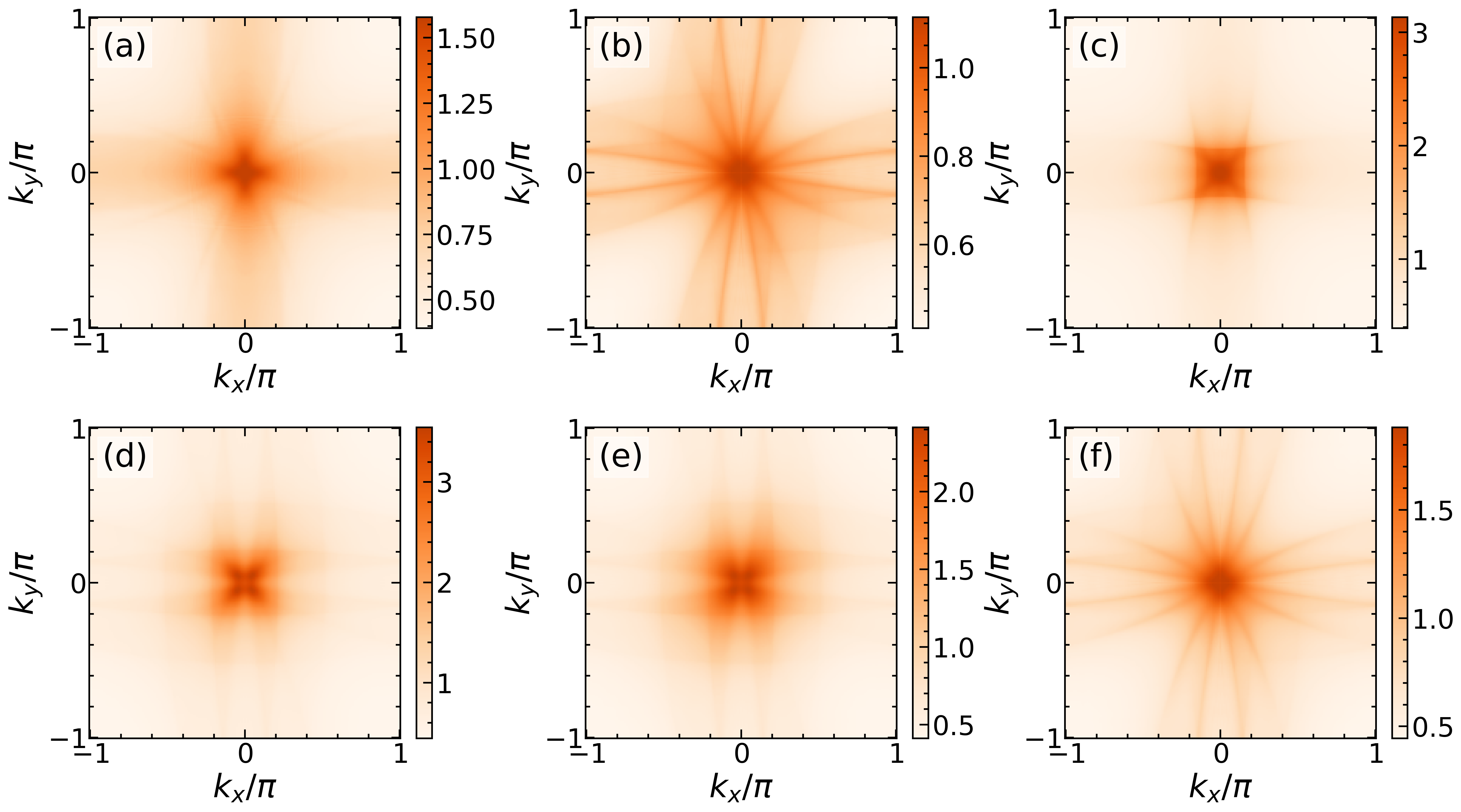}
\caption{Momentum-space distribution of RPA susceptibilities $\chi^{\text{RPA}}_{\alpha}(\bm{k})$.
(a-f) Calculated susceptibility $\chi^{\text{RPA}}_{1 \to 6}(\bm{k})$ for $U=1$ and $J_H=0.2$. The ${\cal O}_4$ phase (panel d) exhibits the strongest susceptibility with pronounced peaks at the $\Gamma$ point (slightly exceeding that of the ${\cal O}_3$ phase in panel (c).
All calculations use the same parameters as in Fig.~\ref{Lattice}(c).
}
\label{allkO5}
\end{figure}

We now introduce the interactions in Eq.~\eqref{Hamiltonian} to determine whether ${\cal O}_1$ and ${\cal O}_3$ remain the leading instabilities in the interacting system. Within the RPA framework, repulsive onsite Hubbard interactions suppress charge susceptibility while enhancing spin susceptibility~\cite{scalapino1986d}. Specifically, the RPA-renormalized static susceptibility for the ${\cal O}_\alpha$ order takes the form
\begin{align} \label{chiO_RPA}
\chi_{\alpha}^{\text{RPA}}(\bm{k}) &= \frac{1}{2} \sum_{l_1l_2l_3l_4}  [ \bar{\cal O}_\alpha ]_{l_1l_2} [  \bar{\cal O}_\alpha ]_{l_3l_4} [\chi^{\text{RPA}}_{\text{spin}}]^{l_1l_2}_{l_3l_4} ,  
\end{align}
where $\chi^{\text{RPA}}_{\text{spin}} \equiv \chi^{\text{RPA}}_{\text{spin}}(\bm{k},i\omega_n = 0)$ denotes the spin susceptibility tensor. It is obtained through the Dyson equation, $\chi^{\text{RPA}}_{\text{spin}}(\bm{k},0) = [I - \chi^{(0)}(\bm{k},0)U^{(s)}]^{-1}\chi^{(0)}(\bm{k},0)$, where $I$ denotes the identity matrix and $U^{(s)}$ the interaction matrix in the spin channel [see Methods]. Then, Eq.~\eqref{chiO_RPA} allows us to compare the susceptibility channels for different competing orders. In general, the spin susceptibility tensor $[\chi^{\text{RPA}}_{\text{spin}}(\bm{k},0)]^{l_1l_2}_{l_3l_4}$ diverges at a critical interaction strength $U_c$ and at an ordering wavevector $\bm{Q}$. Here, $U_c$ determines the magnetic/non-magnetic phase boundary and $\bm{Q}$ characterizes the spatial periodicity of the dominant magnetic phase. For $U>U_c$, the system develops long-range magnetic order. The magnetic phase transition is driven by the divergence of $[\chi^{\text{RPA}}_{\text{spin}}]_{l_3l_4}^{l_1l_2}$ at $U_c$, which simultaneously triggers the divergence of $\chi_{\alpha}^{\text{RPA}}$.

We use the shorthand $\chi_{\Gamma,\alpha}^{\text{RPA}} \equiv \chi_{\alpha}^{\text{RPA}}(\Gamma)$ for notational simplicity. We first examine the case where ${\cal O}_3$ is the leading instability at $U=J_H=0$, using the band parameters of Fig.~\ref{Lattice}(c). Based on Eq.~\eqref{chiO_RPA}, Figure~\ref{allkO5} presents the calculated momentum-resolved RPA susceptibility $\chi_{\alpha}^{\text{RPA}}(\bm{k})$ for $U=1$ and $J_H=0.2$. We find that the dominant susceptibility channel, $\chi_{4}^{\text{RPA}}(\bm{k})$, exhibits a pronounced peak at the $\Gamma$ point. This indicates that, after interaction renormalization, the ${\cal O}_4$ order (extrinsic altermagnetism) becomes the leading instability, rather than ${\cal O}_3$. The resulting magnetic phase transition from ${\cal O}_3$ to ${\cal O}_4$ occurs at weak interaction strengths due to the electron filling being near the VHS.

To systematically explore the magnetic phase transitions, we construct interaction phase diagrams in the $U$–$J_H$ plane for systems exhibiting \textit{intrinsic} or \textit{extrinsic} altermagnetism. Our approach involves three steps:  (i) identifying parameter sets that maximize $\chi_{\alpha}^{\text{RPA}}(\bm{k})$ at the $\Gamma$-point for the target ${\cal O}_\alpha$ order, (ii) determining the critical interaction strength $U_c$; and (iii) identifying the dominant magnetic orders near criticality. Specific examples are presented in the following three cases.

\begin{figure}[!htbp]
\centering
\includegraphics[width=0.48\textwidth]{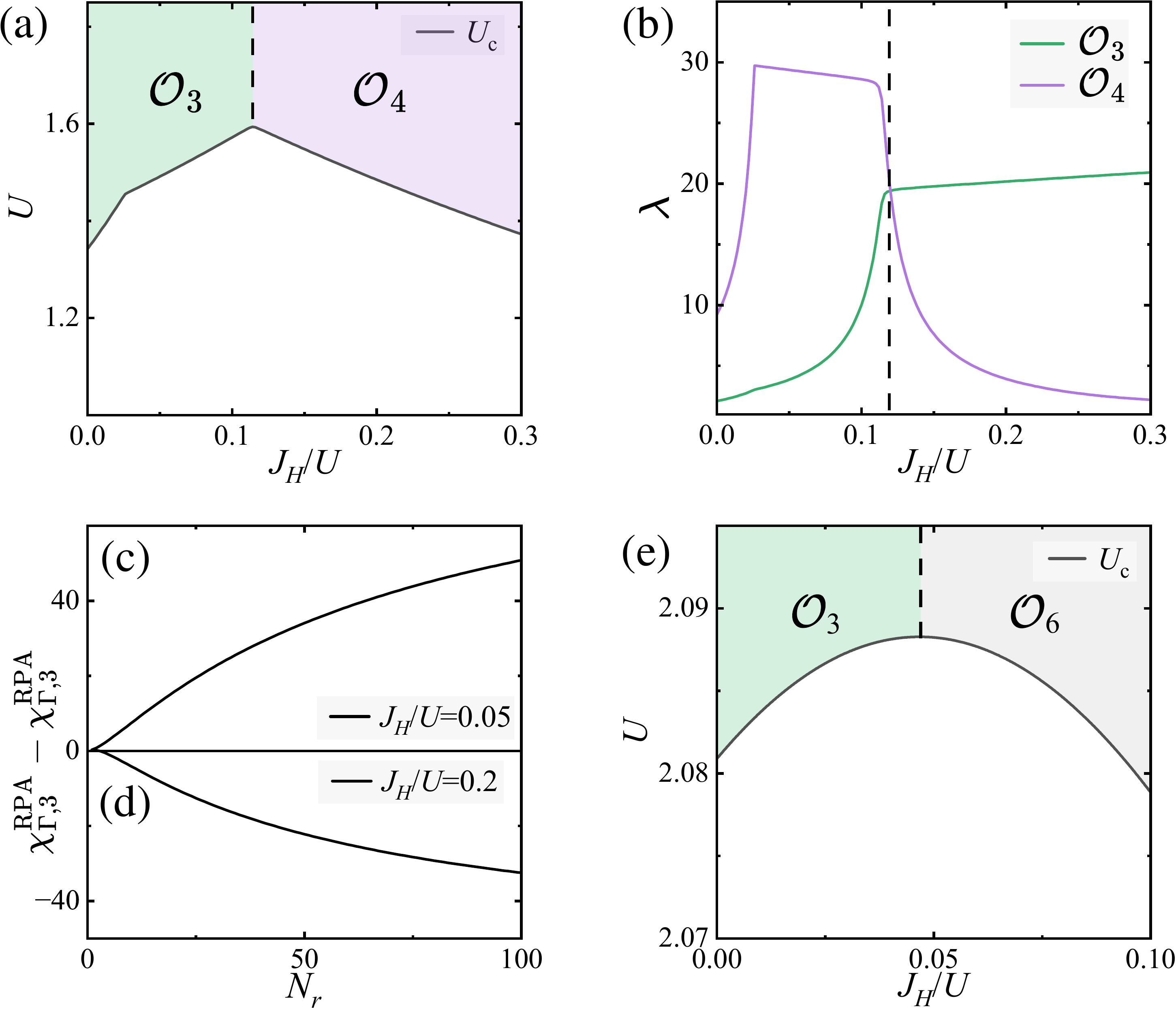}
\caption{The phase diagram with $d$-wave extrinsic altermagnetic phase [Case I].
(a) The $U$–$J_H$ phase diagram for parameters used in Fig.~\ref{Lattice}(c) shows: extrinsic $d$-wave altermagnetic phase (${\cal O}_4$, purple), ferrimagnetic phase (${\cal O}_3$, green), and non-magnetic phase (white). The magnetic-to-non-magnetic boundary is marked by the critical $U_c$ (solid black curve); the dashed line at $J_H/U=0.114$ separates the $\mathcal{O}_3$ and $\mathcal{O}_4$ phases.
(b) The two largest eigenvalues of the susceptibility matrix correspond to the ${\cal O}_3$ and ${\cal O}_4$ phases.
(c,d) Divergence of the susceptibility difference ${\chi}^{\text{RPA}}_{\Gamma,1}-{\chi}^{\text{RPA}}_{\Gamma,6}$ as $U$ approaches $U_c$, with $U=(1-1/N_r)U_c$ and $N_r$ running from $0$ to $200$. Results are shown for $J_H/U=0.05$ (c) and $J_H/U=0.2$ (d).
All calculations employ the band parameters of  Fig.~\ref{Lattice}(c). (e) The transition between ${\cal O}_3$ and ${\cal O}_6$ occurs at $J_H/U=0.047$. Parameters: $\{ t_0,t_1,t_2,t_3,t_5,\mu\} = \{1,0.42,0.08,0.7,0.49,-1\}$.
}
\label{O5}
\end{figure}

\noindent{\bf Case I: phase diagram with $d$-wave extrinsic altermagnetism.}
In our previous calculations, we set $U=1$ and $J_H=0.2$ for Fig.~\ref{allkO5}. We then construct the $U$-$J_H$ phase diagram in Fig.~\ref{O5}(a), which shows a transition from ${\cal O}_3$ at low $J_H$ to ${\cal O}_4$ above a critical ratio $J_H/U=0.114$. The solid black curve marks the boundary between non-magnetic and magnetic states, tracing $U_c$ as a function of $J_H/U$. We emphasize that the ${\cal O}_4$ phase is a $d$-wave extrinsic altermagnet, while the ${\cal O}_3$ phase is a spin-split ferrimagnet. These two orders are unambiguously distinguished through two methods: (i) the eigen-analysis of the spin susceptibility matrix $[\chi^{\text{RPA}}_{\text{spin}}(\Gamma,0)]^{l_1l_1}_{l_3l_3}$ yields the two dominant eigenvalues $\lambda_{{\cal O}_3}$ and $\lambda_{{\cal O}_4}$, whose difference change sign at $J_H/U=0.114$ [Fig.~\ref{O5}(b)]; and (ii) the comparison of divergence rates via the susceptibility difference $\chi_{\Gamma,3}^{\text{RPA}} - \chi_{\Gamma,4}^{\text{RPA}}$ [Figs.~\ref{O5}(c) and (d)]. Applying the same analysis procedure to the parameter set $\{ t_0,t_1,t_2,t_3,t_5,\mu \} = \{1,0.42,0.08,0.7,0.49,-1\}$, we find that increasing $J_H$ drives a transition from ${\cal O}_3$ into a ferromagnetic phase (${\cal O}_6$, gray) [Fig.~\ref{O5}(e)]. Between the two methods, the second approach—relying on the susceptibility difference—is computationally simpler. Further technical details are provided in the Methods section.

\begin{figure}[t]
\centering
\includegraphics[width=0.48\textwidth]{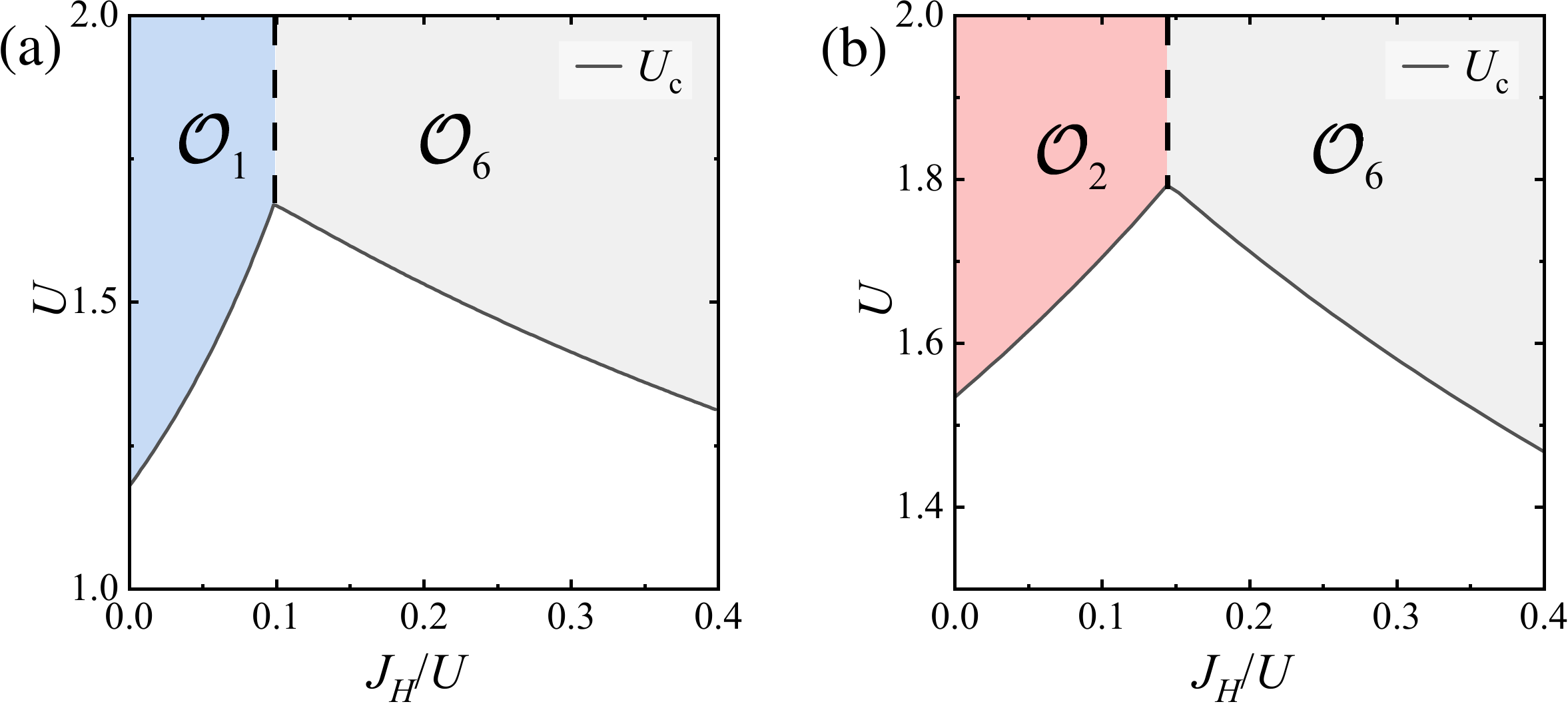}
\caption{The phase diagram with $d$-wave intrinsic altermagnetic phase [Case II].
(a) The phase diagram for parameters used in Fig.~\ref{Lattice}(b) shows: intrinsic $d$-wave altermagnetic phase (${\cal O}_1$, blue), ferromagnetic phase (${\cal O}_6$, gray), and non-magnetic phase (white). The dashed line separates the ${\cal O}_1$ and ${\cal O}_6$ phases at $J_H/U=0.098$. The parameters are the same as Fig.~\ref{Lattice}(b).
(b) Similarly, using another set of band parameters leads to the other intrinsic altermagnetic phase (${\cal O}_2$, red). The phase boundary between $\mathcal{O}_2$ and $\mathcal{O}_6$ is $J_H/U = 0.144$. Parameters: $\{t_0,t_1,t_2,t_3,t_5,\mu \}=\{ 1,0.25,0.25,0.17,0.07,-1\}$.
}
\label{O4}
\end{figure}

\noindent{\bf Case II: phase diagram with $d$-wave intrinsic altermagnetism.}
Using the band parameters in Fig.~\ref{Lattice}(b), where the inter‑orbital hopping parameters $t_3$ and $t_5$ are enhanced relative to those in Fig.~\ref{Lattice}(a) [{\bf Case I}], we obtain the full phase diagram in the $U$–$J_H/U$ plane shown in Fig.~\ref{O4}(a). The phase diagram consists of two distinct magnetic states: $d$-wave intrinsic altermagnetism (${\cal O}_1$, blue), and ferromagnetism (${\cal O}_6$, gray). Hund’s coupling naturally favors parallel spin alignment across orbitals, which stabilizes the ${\cal O}_6$ phase at larger $J_H$. Moreover, for the parameter set $\{t_0,t_1,t_2,t_3,t_5,\mu \}=\{ 1,0.25,0.25,0.17,0.07,-1\}$, we identify an additional intrinsic altermagnetic phase $\mathcal{O}_2$ (red), which emerges over a broader range of $J_H$ [Fig.~\ref{O4}(b)]. Further details on the modified Stoner ferromagnetism and the altermagnetic fluctuation–induced spin-triplet superconductivity can be found in Refs.~\cite{lu2025breakdown,lu2025inter}.

\begin{figure}[b]
\centering
\includegraphics[width=0.48\textwidth]{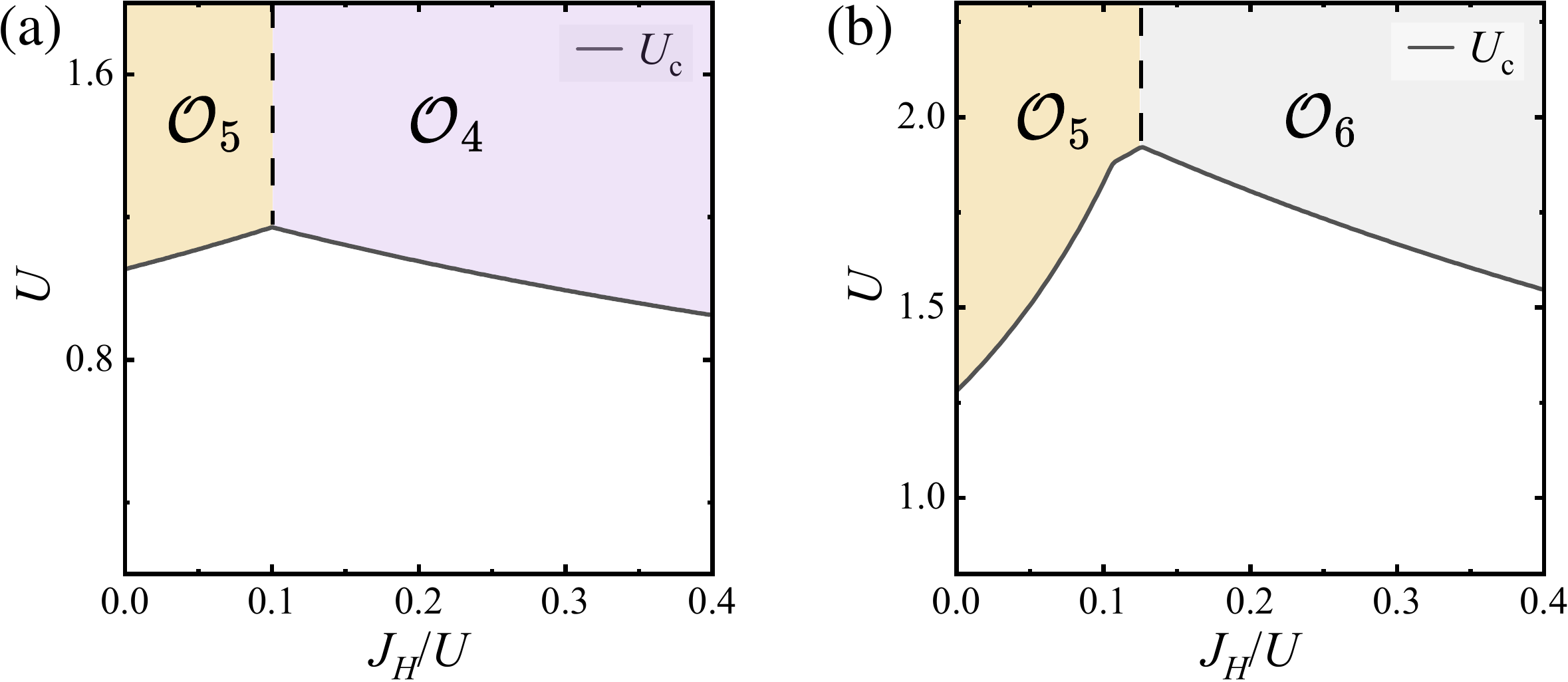}
\caption{The phase diagram with $g$-wave extrinsic altermagnetic phase [Case III].
(a) Phase diagram includes extrinsic $g$-wave altermagnetic phase (${\cal O}_5$, brown), extrinsic $d$-wave altermagnetic phase (${\cal O}_4$, purple), and non-magnetic phase (white). The transition between ${\cal O}_5$ and ${\cal O}_4$ occurs at $J_H/U=0.1$. Parameters: $\{ t_0,t_1,t_2,t_3,t_5,\mu\} = \{1,0.1,0.05,0.14,0.2,-0.26\}$.
(b) shows the transition from ${\cal O}_5$ to ${\cal O}_6$ at $J_H/U=0.126$. Parameters: $\{ t_0,t_1,t_2,t_3,t_5,\mu \} = \{1,0.1,0.05,0.36,0.32,-0.08 \}$.
}
\label{S3}
\end{figure}

\noindent{\bf Case III: phase diagram with $g$-wave extrinsic altermagnetism.}
We now employ a different parameter set $\{ t_0,t_1,t_2,t_3,t_5,\mu\} = \{1,0.1,0.05,0.14,0.2,-0.26\}$, which maintains filling near the VHSs but reduces $t_3$ and enhances $t_5$ relative to the {\bf Case I}. This leads to a distinct $J_H$-driven transition between two extrinsic altermagnetic phases: a $g$-wave phase (${\cal O}_5$, brown) at low $J_H$ and a $d$-wave phase (${\cal O}_4$, purple) at large $J_H$ [Fig.~\ref{S3}(a)], in qualitative agreement with mean‑field results~\cite{wang2024submit}. Moreover, a magnetic phase transition from ${\cal O}_5$ to a ferromagnetic phase (${\cal O}_6$, gray) is also observed in Fig.~\ref{S3}(b), where we use another parameter set, $\{ t_0,t_1,t_2,t_3,t_5,\mu \} = \{1,0.1,0.05,0.36,0.32,-0.08 \}$.

\begin{figure}[t]
\centering
\includegraphics[width=0.4\textwidth]{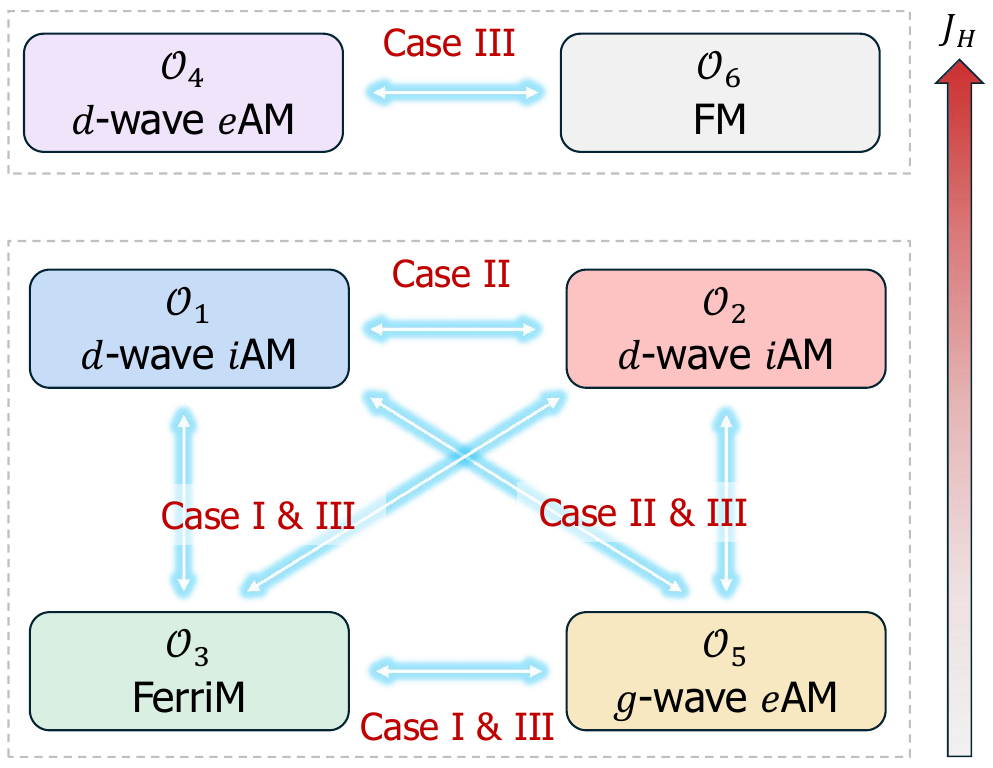}
\caption{A summary of this work: tunable magnetic phase transitions for the six $\mathbf{Q}=\bm{0}$ orders ${\cal O}_{1\to6}$ [see Table~\ref{table2}].
}
\label{fig6}
\end{figure}

\vspace{0.5\baselineskip}
\noindent{\bf Discussions and conclusion}\\
\noindent{\bf Tunable magnetic phase transitions.}
The three cases presented above demonstrate that the magnetic phase transitions are highly tunable, governed by both the kinetic energy (hopping parameters) and the interaction strengths ($U$ and $J_H$). Here, we summarize all possible transitions among the ${\cal O}_{1}$–${\cal O}_{6}$ phases in a two‑orbital system [Fig.~\ref{fig6}]. Larger Hund’s coupling $J_H$ generally favors either the $d$‑wave extrinsic altermagnet ${\cal O}_4$ or the ferromagnetic phase ${\cal O}_6$, whereas reducing $J_H$ can drive the system toward the remaining four phases (${\cal O}_{1,2,3,5}$). Transitions among these four are further mediated by variations in the hopping parameters. For instance, comparing Fig.~\ref{O4}(b) [{\bf Case II}] with Fig.~\ref{S3}(b) [{\bf Case III}] shows that a transition from the $d$‑wave intrinsic altermagnet ${\cal O}_1$ to the $g$‑wave extrinsic altermagnet ${\cal O}_5$ can be realized by decreasing $t_{1,2}$ while increasing $t_{3,5}$. More details are provided in Sec.~D of SM. Although a full $t$–$U$–$J_H$ phase diagram is beyond the scope of this work, the results presented here serve as essential building blocks for understanding the rich landscape of VHS‑induced multiple magnetic phase transitions.

\vspace{0.5\baselineskip}
\noindent{\bf Role of VHS.}
The $\bm{Q}=\bm{0}$ orders emerge precisely as the chemical potential approaches the VHS, highlighting its crucial role. Since $\bm{Q}=\bm{0}$ is a necessary requirement for both \textit{extrinsic} and \textit{intrinsic} altermagnetism, we examine the VHS influence below. In Eq.~\eqref{chi00}, we note that while the normalized eigenvector product contribution remains finite, the Lindhard function term $[\eta_F(\varepsilon^{\beta}_{\bm{k}_1+\bm{Q}})-\eta_F(\varepsilon^{\alpha}_{\bm{k}_1})]/[\varepsilon^{\alpha}_{\bm{k}_1}- \varepsilon^{\beta}_{\bm{k}_1+\bm{Q}}]$ can diverge. At low temperature, this divergence at $\bm{Q} = \bm{0}$ requires three simultaneous conditions: (1) a finite numerator, requiring that the $\alpha$ and $\beta$ bands to be on opposite sides of the Fermi level (one occupied, one unoccupied); (2) a vanishing denominator, demanding both band energies approach the Fermi level; and (3) a sufficiently high density of states (DOS) at $\bm{k}_1$-points where conditions (1) and (2) are simultaneously satisfied, ensuring the integrated susceptibility yields a strong divergence. Therefore, the condition (3) is automatically satisfied when the Fermi level is tuned near a VHS at high-symmetry points, implying the band structure must exhibit a Dirac semi-metal character (e.g.,~$\epsilon_\alpha \sim k^2$ and $\epsilon_\beta \sim -k^2$). This is directly confirmed by Figs.~\ref{Lattice}(b) and (c).

We next show that these conditions maximize the susceptibility at $\bm{Q} = \bm{0}$. As the Fermi level approaches a VHS, a set of high-DOS ``hot spots'' emerges in momentum space. To identify the dominant instability, it suffices to examine the nesting vectors connecting these hot spots, which include both $\bm{Q} \neq \bm{0}$ (inter-VHS) and $\bm{Q} = \bm{0}$ (intra-VHS). When the VHS is located near a single high-symmetry point (e.g., $\bf{\Gamma}$, $\textbf{M}$, $\textbf{X}$, $\dots$), $\bm{Q} = \bm{0}$ is the only nesting vector capable of linking them. Moreover, the band degeneracy at VHS points is typically symmetry-protected, the $\bm{Q} = \bm{0}$ order inherits this protection, enhancing its stability against perturbations.

In the general case where multiple VHSs lie away from high-symmetry points, both inter-VHS and intra-VHS nesting contributions coexist. The intra-VHS typically dominates in the Lindhard function part. For instance, consider two VHSs located at $\bm{q}_1$ and $\bm{q}_2$ with semi-metal dispersions $\epsilon(\bm{k}-\bm{q}_1) = \pm t_1 k^2$ and $\epsilon(\bm{k}-\bm{q}_2) = \pm t_2 k^2$, respectively. The intra-VHS contribution to the Lindhard function scales as $1/(2t_1k^2) + 1/(2t_2k^2)$, while the inter-VHS term varies as $2/(t_1k^2+t_2k^2)$. Consequently, as $k\to 0$, the intra-VHS divergence rate $1/2t_1 + 1/2t_2$ exceeds the inter-VHS rate $2/(t_1+t_2)$ by the harmonic-arithmetic mean inequality. When additional VHSs are present, inter-VHS nesting produces multiple nonzero $\bm{Q}$ vectors that inherently access reduced phase space compared to the intra-VHS case. Therefore, in semi-metal systems with Fermi levels near VHSs, the susceptibility consistently peaks at $\bm{Q}=\bm{0}$. Furthermore, slightly shifting the filling away from the VHSs causes the RPA susceptibilities to rapidly shift their divergence to nonzero $\bm{Q}$, as detailed in the Sec.~C of SM.

\vspace{0.5\baselineskip}
\noindent{\bf Possible candidates.}
The above argument provides a practical guidance for realizing altermagnetism in multi-orbital quantum materials. Promising candidate systems include transition metal oxides where orbital order coexists with antiferromagnetism, such as: $\alpha-$Sr$_2$CrO$_4$ with a 3$d^2$ electronic configuration in the Cr$^{4+}$ state~\cite{pandey2021origin,lee2022ultrafast}; SrRuO$_3$ thin films with SrO termination~\cite{autieri2016antiferromagnetic}; various vanadium-based oxides such as V$_2$O$_3$~\cite{shiina2001atomic}, ZnV$_2$O$_4$~\cite{maitra2007orbital}; and a class of manganites exhibiting robust orbital ordering, including Mn$_2$OBO$_3$~\cite{goff2004spin}, Pr$_{1-x}$Ca$_x$MnO$_3$~\cite{zimmermann1999interplay}, La$_{0.5}$Sr$_{1.5}$MnO$_4$~\cite{ehrke2011photoinduced}, and La$_{0.5}$Ca$_{0.5}$MnO$_3$~\cite{radaelli1997charge}. These orbital-active magnetic materials may provide a platform to explore various kinds of altermagnetic phases, including intrinsic altermagnetism.

\vspace{0.5\baselineskip}
\noindent{\bf Conclusion.}
This work establishes a universal mechanism for realizing intrinsic altermagnetism in correlated multi-orbital systems, driven by orbital antiferromagnetism with antiparallel spins locked to distinct orbitals. Using symmetry analysis and random-phase-approximation calculations, we identify the essential criteria: suppressed Hund's coupling, dominant inter-orbital hopping, and electron filling near Van Hove singularities from quadratic band touching. Phase diagrams in the Hubbard $U$–$J_H$ plane reveal rich magnetic competition, including phase transitions between intrinsic altermagnetism, extrinsic altermagnetism, ferromagnetism, and ferrimagnetism. Our results provide a clear theoretical pathway for designing altermagnetic states in correlated materials through manipulation of electronic structure and correlations.

\vspace{\baselineskip}
\noindent{\bf Methods} \\
\noindent{\bf The Multi-Orbital RPA Approach}\\
The explicit formulism of $\chi^{(0)}$ in the momentum-frequency space is,
\begin{align}\label{chi00}
\begin{split}
&[\chi^{(0)}(\bm{k},i \omega)]^{l_1l_2}_{l_3l_4} \equiv \frac{1}{N} \sum_{\bm{k}_1\alpha \beta}  [\xi^{\alpha}_{l1}(\bm{k}_1)]^\ast \xi^{\beta}_{l2}(\bm{k}_1+\bm{k}) \times \\
&\qquad\quad [\xi^{\beta}_{l3}(\bm{k}_1+\bm{k})]^\ast  \xi^{\alpha}_{l4}(\bm{k}_1)\frac{\eta_F(\varepsilon^{\beta}_{\bm{k}_1+\bm{k}})-\eta_F(\varepsilon^{\alpha}_{\bm{k}_1})}{i\omega+\varepsilon^{\alpha}_{\bm{k}_1}- \varepsilon^{\beta}_{\bm{k}_1+\bm{k}}},
\end{split}
\end{align}
where $\alpha, \beta \in \left \{ 1, 2, 3, 4 \right \}$ are band indices, $\varepsilon^{\alpha}_{\bm{k}}$ and $\xi^{\alpha}(\bm{k})$ are the $\alpha$-th eigenvalue and eigenvector of $H_{tb}(\bm{k})$ given by Eq.~\eqref{Hamiltonian_tb}, respectively, and $\eta _F$ is the Fermi-Dirac distribution function. In our calculations, a lattice size of $N=200\times 200$ is employed. To enhance the smoothness of the results and reduce the influence of finite-size effects, we set the temperature to $0.005$. This temperature is sufficiently low to ensure that our conclusions remain unaffected by thermal fluctuations.

\begin{figure}[htbp]
\centering
\includegraphics[width=0.46\textwidth]{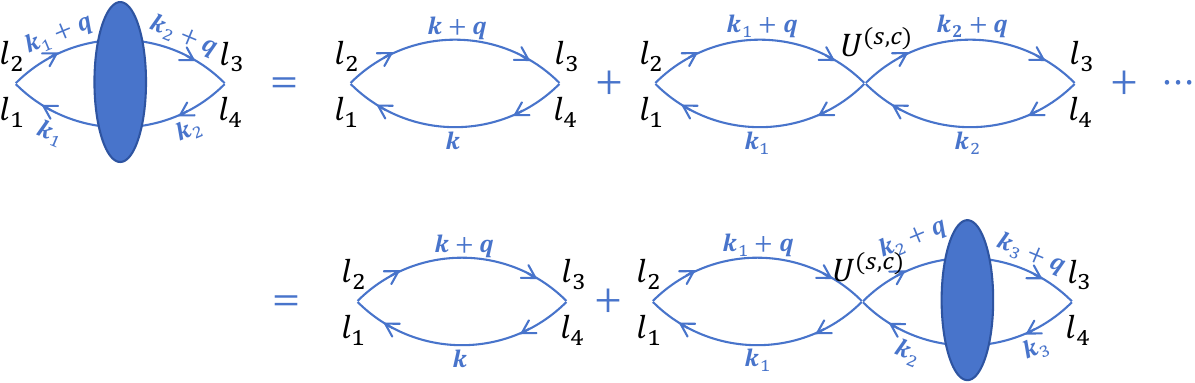}
\caption{Feyman’s diagram for the renormalized spin/charge susceptibility in the RPA level.}
\label{figs1}
\end{figure}

When the interactions in Eq.~\eqref{Hamiltonian} are turned on, the spin and charge susceptibilities are defined as,
\begin{align}\label{chis}
\begin{split}
&[\chi_{\text{spin}}(\bm{k},\tau)]^{l_1l_2}_{l_3l_4}
\equiv \frac{1}{2N} \sum_{\bm{k}_1\bm{k}_2} \sum_{s s^{\prime}} ss^{\prime} \langle T_{\tau} c^{\dagger}_{l_1,s}(\bm{k}_1,\tau)  \\
& \quad\quad  c_{l_2,s}(\bm{k}_1+\bm{k},\tau)c^{\dagger}_{l_3,s^{\prime}}(\bm{k}_2+\bm{k},0)c_{l_4,s^{\prime}}(\bm{k}_2,0)\rangle,
\end{split}
\end{align}
and 
\begin{align}\label{chisc}
\begin{split}
&[\chi_{\text{charge}}(\bm{k},\tau)]^{l_1l_2}_{l_3l_4} \equiv \frac{1}{2N} \sum_{\bm{k}_1 \bm{k}_2} \sum_{s s^{\prime}} \langle T_{\tau} c^{\dagger}_{l_1,s}(\bm{k}_1,\tau) \\
&\quad\quad c_{l_2,s}(\bm{k}_1+\bm{k},\tau)c^{\dagger}_{l_3,s^{\prime}}(\bm{k}_2+\bm{k},0)c_{l_4,s^{\prime}}(\bm{k}_2,0) \rangle, 
\end{split}
\end{align}
where $\left \langle \dots  \right \rangle $ denotes the thermal average for the interacting system, and $s,s'=\pm$ label the spin indices. In the non-interacting limit ($U = V = J_H = 0$), the spin and charge susceptibilities coincide with the bare susceptibility: $\chi_{\text{spin}}=\chi_{\text{charge}}=\chi^{(0)}$. When interactions are introduced, the random-phase approximation yields the renormalized susceptibilities,
\begin{align}\label{chisc}
\begin{split}
\chi_{\text{spin}}^{\text{RPA}}(\bm{k},i\omega) &=  [I - \chi^{(0)}(\bm{k},i\omega)U^{(s)} ]^{-1} \chi^{(0)}(\bm{k},i\omega), \\
\chi_{\text{charge}}^{\text{RPA}}(\bm{k},i\omega) &=  [I + \chi^{(0)}(\bm{k},i\omega)U^{(c)}]^{-1} \chi^{(0)}(\bm{k},i\omega), 
\end{split}
\end{align}
where the interaction matrices $U^{(s)}$ and $U^{(c)}$ have identical dimensions and are corresponding to the spin and charge channels respectively. For calculation simplicity, we express the susceptibility tensor in matrix form, where rows correspond to orbital indices $l_1l_2$ and columns to $l_3l_4$. Figure~\ref{figs1} shows the standard RPA-level Feynman diagrams for both spin and charge susceptibilities. We label the sublattice-orbital basis $\{ \vert A,d_{xz} \rangle, \vert A,d_{yz} \rangle, \vert B,d_{xz} \rangle, \vert B,d_{yz} \rangle \}$ as $\{1, 2, 3, 4\}$. The non-zero elements of the interaction tensor $[U^{(s,c)}]^{l_1l_2}_{l_3l_4}$ are then given by:
\begin{align} \label{Usc}
&[U^{(s)}]^{11}_{11} = [U^{(s)}]^{22}_{22} = [U^{(s)}]^{33}_{33} =[U^{(s)}]^{44}_{44} = U, \\
&[U^{(s)}]^{11}_{22} = [U^{(s)}]^{22}_{11} = [U^{(s)}]^{33}_{44} = [U^{(s)}]^{44}_{33} = J_H, \nonumber \\
&[U^{(s)}]^{12}_{12} = [U^{(s)}]^{21}_{21} = [U^{(s)}]^{34}_{34} = [U^{(s)}]^{43}_{43} = J_H, \nonumber \\
&[U^{(s)}]^{12}_{21} = [U^{(s)}]^{21}_{12} = [U^{(s)}]^{34}_{43} = [U^{(s)}]^{43}_{34} = V, \nonumber
\end{align}
and
\begin{align}
&[U^{(c)}]^{11}_{11} = [U^{(c)}]^{22}_{22} = [U^{(c)}]^{33}_{33} = [U^{(c)}]^{44}_{44} = U, \\
&[U^{(c)}]^{11}_{22} = [U^{(c)}]^{22}_{11} = [U^{(c)}]^{33}_{44} = [U^{(c)}]^{44}_{33} = 2V-J_H, 
\nonumber \\
&[U^{(c)}]^{12}_{12} = [U^{(c)}]^{21}_{21} = [U^{(c)}]^{34}_{34} = [U^{(c)}]^{43}_{43} = J_H, \nonumber \\
&[U^{(c)}]^{12}_{21} = [U^{(c)}]^{21}_{12} = [U^{(c)}]^{34}_{43} = [U^{(c)}]^{43}_{34} = 2J_H-V. \nonumber
\end{align}

\vspace{0.5\baselineskip}
\noindent{\bf Magnetic Order from Susceptibility Analysis}\\
To analyze the magnetic order, we first introduce the Matsubara Green’s function in momentum space,
\begin{align}
\begin{split}
{\cal G}_{l_1,l_3}(\bm{k},\tau)\equiv \left \langle {\cal T}_{\tau} S^{z}_{l_1}(\bm{k},\tau) S^{z}_{l_3}(-\bm{k},0) \right \rangle.
\end{split}
\end{align}
Here $\tau$ is the imaginary time and $\cal T_{\tau}$ represents the time-ordered product of operators and ${l_1}, {l_3}\in\left \{ 1, 2, 3, 4 \right \} $ are sublattice-orbital indices. Its Fourier transform gives the static spin susceptibility $\chi^{(s)}_{l_1,l_3}(\bm{k})$

\begin{align}
\begin{split}
&\chi^{(s)}_{l_1,l_3}(\bm{k})\equiv {\cal G}_{l_1,l_3}(\bm{k},i\omega_n=0),\\
&{\cal G}_{l_1,l_3}(\bm{k},i\omega_n)=\int_{0 }^{\beta} d\tau {\cal G}_{l_1,l_3}(\bm{k},\tau)e^{i\omega_n\tau}.
\end{split}
\end{align}
Where $\beta$ is the inverse temperature and $\omega_n=\frac{2\pi n}{\beta } $ is the Matsubara frequency. In the RPA-level $\chi^{(s)}_{l_1,l_3}(\bm{k})\equiv [\chi^{\text{RPA}}_{\text{spin}}(\bm{k})]^{l_1,l_1}_{l_3,l_3}$. From linear response theory, as the interaction strength $U\to U_c^{-}$, the spin susceptibility matrix $[\chi^{\text{RPA}}_{\text{spin}}(\bm{Q})]^{l_1,l_1}_{l_3,l_3}$ diverges at the ordering wave vector $\bm{Q}$. Consequently, the system undergoes a magnetic transition, and the resulting magnetic order is determined by the eigenvector $\bm{Q}$ associated with the largest eigenvalue of the susceptibility matrix.

However, when the magnetic order parameter has a more complex structure, this approach requires an additional eigenvector analysis to obtain the correct result [see next section]. In this case, our second criterion for identifying the magnetic order may be more intuitive and reliable. Specifically, as $U \to U_c^{-}$, one can directly compare the magnitudes of the susceptibilities corresponding to different magnetic channels at the divergent wave vector $\bm{Q}$ [see Eq.~\eqref{chiO_RPA}], and the system will enter the channel with the most strongly diverging susceptibility.

\vspace{0.5\baselineskip}
\noindent{\bf The Eigen-vector Analysis from Susceptibility}\\
In the main text, we employed a key result: the eigenvector associated with the largest eigenvalue of the static spin susceptibility matrix, $[\chi_{\text{spin}}^{\text{RPA}}(\bm{k},0)]_{l_3l_3}^{l_1l_1}$, directly corresponds to the order parameter of the magnetic phase. To clarify, the orbital indices $l_1=l_2$ and $l_3=l_4$ for the general spin susceptibility $[\chi_{\text{spin}}^{\text{RPA}}(\bm{k},0)]_{l_3l_4}^{l_1l_2}$ indicate that this matrix describes spin-spin correlations between these two orbitals ($l_1$ and $l_3$). In our work, $l_1$ represents the basis $\{ \vert A, d_{xz} \rangle, \vert B, d_{xz} \rangle, \vert A, d_{yz} \rangle, \vert B, d_{yz} \rangle \}$. Specifically, the susceptibility matrix captures the linear response of spins in orbital $l_1$ to perturbations in orbital $l_3$ (and vice versa), formally defined as:
\begin{align}
[\chi_{\text{spin}}&(\bm{k},\tau)]^{l_1l_1}_{l_3l_3}
\equiv \frac{1}{2N} \sum_{\bm{k}_1 \bm{k}_2} \sum_{s s^{\prime}} ss^{\prime} \langle T_{\tau} c^{\dagger}_{l_1,s}(\bm{k}_1,\tau)  \\
& \times c_{l_2,s}(\bm{k}_1+\bm{k},\tau)c^{\dagger}_{l_3,s^{\prime}}(\bm{k}_2+\bm{k},0)c_{l_4,s^{\prime}}(\bm{k}_2,0)\rangle, \nonumber
\end{align}
where $\left \langle \dots  \right \rangle $ denotes the thermal average for the interacting system, and $s,s'=\pm$ label the spin indices. To account for electron-electron interactions, we compute the spin susceptibility within the random phase approximation (RPA), denoted as $[\chi_{\text{spin}}^{\text{RPA}}(\bm{k},i\omega_n = 0)]^{l_1l_1}_{l_3l_3}$ in the static limit. Note that $[\chi_{\text{spin}}^{\text{RPA}}(\bm{k},i\omega_n = 0)]^{l_1l_1}_{l_3l_3}$ is a positive definite matrix. The free energy change due to magnetic ordering can be expressed as
\begin{align}
\Delta F \propto \sum_{\bm{k}}\sum_{l_1l_3} M_{l_1}(\bm{k})  [\chi_{\text{spin}}^{\text{RPA}}(\bm{k},0)]^{l_1l_1}_{l_3l_3} M_{l_3}(\bm{k}),
\end{align}
where $M_{l_1}(\bm{k})$ denotes the spin density on the $l_1$ orbital. We diagonalize the static spin susceptibility matrix,
\begin{align}
[\chi_{\text{spin}}^{\text{RPA}}(\bm{k},0)]^{l_1l_1}_{l_3l_3} = \sum_{i=1}^{4} \lambda_i \vert \lambda_i \rangle \langle \lambda_i \vert,    
\end{align}
where $\lambda_i \geq 0$ are the eigenvalues. Let $\lambda_{\text{max}}$ be the largest eigenvalue, and decompose the spin density as $\Vec{M}(\bm{k}) = M_0 \hat{M}(\bm{k})$, where $M_0$ is the amplitude and $\hat{M}(\bm{k})$ is a unit vector. The free energy change is then bounded by: $\Delta F \leq M_0^2 \lambda_{\text{max}}$, with equality achieved when $\hat{M}(\bm{k})$ aligns with the eigenvector $\vert \lambda_{\text{max}} \rangle$,
\begin{align}
\hat{M}(\bm{k}) = \vert \lambda_{\text{max}} \rangle.    
\end{align}
This indicates that the system develops a dominant magnetic instability when $\lambda_{\text{max}}$ diverges at the critical interaction strength $U\to U_c^{-}$, marking the transition to long-range magnetic order. Consequently, the corresponding eigenvector $\vert \lambda_{\text{max}} \rangle$ directly reveals the symmetry and orbital structure of this emergent magnetic phase.

In this work, we investigate six distinct magnetic orders characterized by zero momentum $\bm{Q}=0$. These orders are defined by the following operators: ${\cal O}_{1\to6} = \{ \tau_z \sigma_0, \tau_x\sigma_0, \tau_z\sigma_z, \tau_0\sigma_z, \tau_x\sigma_z, \tau_0\sigma_0 \} \otimes s_z $. For clarity, we focus our analysis on two representative cases:
\begin{itemize}
\item The first diagonal case: $\{ {\cal O}_1, {\cal O}_3, {\cal O}_4, {\cal O}_6 \}$ for the basis $\{ \vert A, d_{xz} \rangle, \vert B, d_{xz} \rangle, \vert A, d_{yz} \rangle, \vert B, d_{yz} \rangle \}$. In this case, we have
\begin{align}
\begin{split}
&{\cal O}_1 = \sigma_0 \tau_z s_z, \;
{\cal O}_3 = \sigma_z \tau_z s_z, \; \\
&{\cal O}_4 = \sigma_z \tau_0 s_z, \;
{\cal O}_6 = \sigma_0 \tau_0 s_z.     
\end{split}
\end{align}
For example, we consider the ${\cal O}_1$-${\cal O}_6$ phase transition presented in Fig.~\ref{O4} in the main text, and use $J_H/U=0.2$ and $U=0.99U_c$. The spin susceptibility matrix $[\chi_{\text{spin}}^{\text{RPA}}(0,0)]^{l_1l_1}_{l_3l_3}$ reads,
\begin{align}
\left(\begin{array}{cccc}
 2.45091 & 0.563459 & 1.41528 & 1.18912 \\
 0.563459 & 2.45091 & 1.18912 & 1.41528 \\
 1.41528 & 1.18912 & 2.45091 & 0.563459 \\
 1.18912 & 1.41528 & 0.563459 & 2.45091 \\
\end{array} \right),
\end{align}
whose eigenvalues and eigenvectors are 
\begin{subequations}
\begin{align}
\lambda_{\text{max}} &= 5.61877, \;\; \vert \lambda_{\text{max}} \rangle = (1,1,1,1)^T/2, \\
\lambda_{2} &= 2.1136, \;\; \vert \lambda_{2} \rangle = (1, -1, 1, -1)^T/2, \\
\lambda_{3} &= 1.66129, \;\; \vert \lambda_{3} \rangle = (1, -1, -1, 1)^T/2, \\
\lambda_{4} &= 0.409962, \;\; \vert \lambda_{4} \rangle = (1, 1, -1, -1)^T/2, 
\end{align}
\end{subequations}

where $\vert \lambda_{\text{max}} \rangle$ represents the order parameter of the ${\cal O}_6$ phase, while the subdominant eigenvector $\vert \lambda_{2} \rangle$ corresponds to the ${\cal O}_1$ phase.
\item The second diagonal case: $\{ {\cal O}_2, {\cal O}_4, {\cal O}_5, {\cal O}_6 \}$ for the basis $\{ \vert A, d_{+} \rangle, \vert B, d_{+} \rangle, \vert A, d_{-} \rangle, \vert B, d_{-} \rangle \}$, where $d_{\pm} = (d_{xz} \pm d_{yz})/\sqrt{2}$. In this case, 
\begin{align}
\begin{split}
&{\cal O}_2 = \sigma_0 \tau_z s_z, \;
{\cal O}_4 = \sigma_z \tau_0 s_z, \;  \\ 
&{\cal O}_5 = \sigma_z \tau_z s_z, \;
{\cal O}_6 = \sigma_0 \tau_0 s_z.      
\end{split}
\end{align}
For example, we consider the ${\cal O}_5$-${\cal O}_4$ phase transition presented in Fig.~\ref{S3} in the following section, and use $J_H/U=0.2$ and $U=0.99U_c$. The spin susceptibility matrix $[\chi_{\text{spin}}^{\text{RPA}}(0,0)]^{l_1l_1}_{l_3l_3} $ reads,
\begin{align}
\left(\begin{array}{cccc}
 8.00491 & 5.43847 & -7.0189 & -4.99188 \\
 5.43847 & 8.00491 & -4.99188 & -7.0189 \\
 -7.0189 & -4.99188 & 8.00491 & 5.43847 \\
 -4.99188 & -7.0189 & 5.43847 & 8.00491 \\
\end{array} \right),
\end{align}
whose eigenvalues and eigenvectors are 
\begin{subequations}
\begin{align}
\lambda_{\text{max}} &= 25.4542, \;\; \vert \lambda_{\text{max}} \rangle = (1,1,-1,-1)^T/2, \\
\lambda_{2} &= 4.59346, \;\; \vert \lambda_{2} \rangle = (1, -1, -1, 1)^T/2, \\
\lambda_{3} &= 1.4326, \;\; \vert \lambda_{3} \rangle = (1, 1, 1, 1)^T/2, \\
\lambda_{4} &= 0.539424, \;\; \vert \lambda_{4} \rangle = (1, -1, 1, -1)^T/2, 
\end{align}
\end{subequations}
where $\vert \lambda_{\text{max}} \rangle$ represents the order parameter of the ${\cal O}_4$ phase, while the subdominant eigenvector $\vert \lambda_{2} \rangle$ corresponds to the ${\cal O}_5$ phase.
\end{itemize}

The eigenvector analysis of the spin susceptibility matrix directly determines the dominant magnetic order. This conclusion aligns with the divergence behavior (i.e.,~pronounced divergence rate) observed in the RPA-calculated susceptibility $\chi_{\Gamma,\alpha}^{\text{RPA}}$. Notably, the $\chi_{\Gamma,\alpha}^{\text{RPA}}$ approach offers a more general framework, as it bypasses the need for basis transformations.

\noindent{\bf Acknowledgments}\\
\noindent We thank C.~Li, H.~K.~Jin, Z.~M.~Pan, M.~Zhang, J.~Kang and F.~Yang for helpful discussions. We thank S.~B.~Zhang for careful reading of the manuscript. 
L.H.H. is supported by National Key R\&D Program of China (Grant No. 2025YFA1411501), the National Natural Science Foundation of China (Grant Nos. 12561160109, 1257040632), the Fundamental Research Funds for the Central Universities (Grant No. 226-2024-00068).
C.L. is supported by the National Natural Science Foundation of China under the Grants No. 12304180.

\bibliography{references}






\end{document}